# Secure Link State Routing for Mobile Ad Hoc Networks

Panagiotis Papadimitratos
*School of Electrical and Computer Engineering*
*Cornell University, Ithaca NY 14853*
papadp@ece.cornell.edu

Zygmunt J. Haas
*School of Electrical and Computer Engineering*
*Cornell University, Ithaca NY 14853*
haas@ece.cornell.edu

**Abstract**

*The secure operation of the routing protocol is one of the major challenges to be met for the proliferation of the Mobile Ad hoc Networking (MANET) paradigm. Nevertheless, security enhancements have been proposed mostly for reactive MANET protocols. The proposed here Secure Link State Routing Protocol (SLSP) provides secure proactive topology discovery, which can be multiply beneficial to the network operation. SLSP can be employed as a stand-alone protocol, or fit naturally into a hybrid routing framework, when combined with a reactive protocol. SLSP is robust against individual attackers, it is capable of adjusting its scope between local and network-wide topology discovery, and it is capable of operating in networks of frequently changing topology and membership.*

## 1. Introduction

The collaborative, self-organizing environment of the *Mobile Ad Hoc Networking (MANET)* technology opens the network to numerous security attacks that can actively disrupt the routing protocol and disable communication. Recently, a number of protocols have been proposed to secure the route discovery process in frequently changing MANET topologies. These protocols are designed to perform route discovery only when a source node needs to route packets to a destination; that is, they are reactive routing protocols [1-3]. Nevertheless, in many cases, proactive discovery of topology can be more efficient; e.g., in networks with low- to medium-mobility, or with high connection rates and frequent communication with a large portion of the network nodes. Furthermore, hybrid routing protocols [4], which are the middle ground, have been shown to be capable of adapting their operation to achieve the best performance under differing operational conditions through locally proactive and globally reactive operation.

In this paper, we study how to provide secure proactive routing and we propose a proactive MANET protocol that secures the discovery and the distribution of link state information across mobile ad hoc domains. Our goal is to provide correct (i.e., factual), up-to-date, and authentic link state information, robust against Byzantine behavior and failures of individual nodes. The choice of a link state protocol provides such robustness, unlike distance vector protocols [5], which can be significantly more affected by a single misbehaving node. Furthermore, the availability of explicit connectivity information, present in link state protocols, has additional benefits: examples include the ability of the source to determine and route simultaneously across multiple routes [6], the utilization of the local topology for efficient dissemination of data [7] or efficient propagation of control traffic [8]. Finally, a wide range of MANET instances is targeted by our design, which avoids restrictive assumptions on the underlying network trust and membership, and does not require specialized node equipment (e.g., GPS or synchronized clocks).

We present here our *Secure Link State Protocol (SLSP)* for mobile ad hoc networks, which is robust against individual attackers. SLSP shares security goals and bears some resemblance to secure link state routing protocols proposed for the "wired" Internet, but, at the same time, it is tailored to the salient features of the MANET paradigm. More specifically, SLSP does rely on the requirements of the robust flooding protocol [9], that is, a central entity to distribute all keys throughout the network and the reliable flooding of link state updates throughout the entire network. SLSP does not seek to synchronize the topology maps across all nodes or to support the full exchange of link state databases [10]. Note that nodes cannot be provided with credentials to prove their authorization to advertise specific routing information [11] due to the continuously changing network connectivity and membership. Finally, the participation of nodes in routing does not stem from their possession of credentials [12], since in MANET, all nodes are expected to equally assist the network operation.

First we present our assumptions and network model, followed by an overview and the definition of SLSP. Next, we discuss a number of relevant issues and conclude by describing related future work.

## 2. SLSP Definition

The *Secure Link State Protocol (SLSP)* for mobile ad hoc networks is responsible for securing the discovery and distribution of link state information. The scope of SLSP may range from a secure neighborhood discovery to a network-wide secure link state protocol. SLSP nodes disseminate their link state updates and maintain topological information for the subset of network nodes within *R* hops, which is termed as their *zone* [4]. Nevertheless, SLSP is a self-contained link state discovery protocol, even though it draws from, and naturally fits within, the concept of hybrid routing.

## 2.1. Assumptions and network model

Each node is equipped with a public/private key pair, namely $E_V$ and $D_V$, and with a single network interface per node within a MANET domain.[1] Key certification can be provided by a coalition of $K$ nodes and the use of threshold cryptography [15,13], the use of local repositories of certificates provided by the network nodes [14], or a distributed instantiation of a *CA* [15].

Nodes are identified by their *IP* addresses, which may be assigned by a variety of schemes, e.g., dynamically or even randomly [16]. Although $E_V$ does not need to be tied to the node's *IP* address, it could be beneficial to use *IP* addresses derived from the nodes' public keys [17]. Nodes are equipped with a one-way or hash function $H$ [18,19] and a public key cryptosystem.

Adversaries may disrupt the protocol operation by exhibiting arbitrary malicious behavior: e.g., replay, forge, corrupt link state updates, try to influence the topology view of benign nodes, or exploit the protocol to mount *Denial of Service (DoS)* attacks.

SLSP is concerned solely with securing the topology discovery; it does not guarantee that adversaries, which complied with its operation during route discovery, would not attempt to disrupt the actual data transmission at a later time. The protection of the data transmission is a distinct problem, which we address in a different publication [6].

## 2.2. Overview

To counter adversaries, SLSP protects link state update (*LSU*) packets from malicious alteration, as they propagate across the network. It disallows advertisements of non-existent, fabricated links, stops nodes from masquerading their peers, strengthens the robustness of neighbor discovery, and thwarts deliberate floods of control traffic that exhausts network and node resources.

To operate efficiently in the absence of a central key management, SLSP provides for each node to distribute its public key to nodes within its zone. Nodes periodically broadcast their certified key, so that the receiving nodes validate their subsequent link state updates. As the network topology changes, nodes learn the keys of nodes that move into their zone, thus keeping track of a relatively limited number of keys at every instance.

SLSP defines a secure neighbor discovery that binds each node *V* to its Medium Access Control (*MAC*) address and its *IP* address, and allows all other nodes within transmission range to identify *V* unambiguously, given that they already have $E_V$.

Nodes advertise the state of their incident links by broadcasting periodically signed link state updates (*LSU*). SLSP restricts the propagation of the *LSU* packets to within the zone of their origin node. Receiving nodes validate the updates, suppress duplicates, and relay previously unseen updates that have not already propagated *R* hops. Link state information acquired from validated *LSU* packets is accepted only if both nodes incident on each link advertise the same state of the link.

## 2.3. Neighbor Discovery

Each node commits its Medium Access Control *(MAC)* address and its *IP* address, the *($MAC_V$, $IP_V$)* pair, to its neighbors by broadcasting signed *hello* messages. Receiving nodes validate the signature and retain the information; in the case of *SUCV* addresses [17] the confirmation for the *IP* address can be done in a memory-less manner.

The proposed binding of the $MAC_V$ strengthens the robustness of our scheme, by disallowing nodes from appearing as multiple ones at the data link layer, and by assisting in protection against flooding DoS attacks.

To achieve these goals, we propose that the *Neighbor Lookup Protocol (NLP)* be an integral part of SLSP. NLP is responsible for the following tasks: (i) maintaining a mapping of *MAC* and *IP* layer addresses of the node's neighbors, (ii) identifying potential discrepancies, such as the use of multiple *IP* addresses by a single data-link interface, and (iii) measuring the rates at which control packets are received from each neighbor, by differentiating the traffic primarily based on *MAC* addresses. The measured rates of incoming control packets are provided to the routing protocol. This way, control traffic originating from nodes that selfishly or maliciously attempt to overload the network can be discarded.

Basically, NLP extracts and retains the 48-bit hardware source address for each received (overheard) frame, along with the encapsulated *IP* address. This requires a simple modification of the device driver [18], so that the data link address is "passed up" to the routing protocol along with each packet. With nodes operating in promiscuous mode, the extraction of such pairs of addresses from all overheard packets leads to a significant reduction in the use of the neighbor discovery and query/reply mechanisms for medium access control address resolution.

Each node updates its neighbor table by retaining both, the data-link and the network interface addresses addresses. The mappings between the two addresses are retained in the table as long as transmissions from the corresponding neighboring nodes are overheard; a *lost neighbor* timeout period[2] is associated with each table entry.

NLP issues a notification to SLSP, according to the content of a received packet, in the event that: (i) a neighbor used an *IP* address different from the address currently recorded in the neighbor table, (ii) two neighbors used the same *IP* address (that is, a packet appears to originate from a node that may have "spoofed" an *IP* address), (iii) a node uses the same medium access control address as the detecting node (in that case, the data link address may be "spoofed"). Upon reception of the notification, the routing protocol discards the packet bearing the address that violated the aforementioned policies.

---

[1] To support operation with multiple interfaces, one key pair should be assigned to each interface.

[2] The *lost neighbor* timeout should be longer than the timeout periods associated with the flushing of routing information (link state, routing table entries), related to the particular neighbor.

### 2.4. Link State Updates

Link state updates are identified by the *IP* address of their originator and a 32-bit sequence number, which provides an ample space of approximately four billion updates. To ensure that the LSU's propagate only within the zone of its origin, i.e., $R$ hops away, the node selects a random number $X$ and calculates a hash chain: $X_i = H^i(X)$, $i=1,...,R$, $H^0(X)=X$. It places $X_R$ and $X_1$ in the *zone_radius* and the *hops_traversed* fields of the LSU header,[3] respectively, and sets *TTL* equal to $R-1$, with $R$ placed in the $R_{LSU}$ field. Finally, a signature is appended, with the header format is shown in Figure 1.

Receiving nodes check if they have the public key of the originating node, unless the key is attached to the *LSU* (see section 2.5 below). For an *LSU* that has already traveled over $i$ hops ($i=R-TTL$), if $i$ is less than the radius of the originating node, the packet is not relayed unless $H^{R-i}(hops\_traversed)$ equals *zone_radius*. Each relaying node sets *hops_traversed* equal to $H(hops\_traversed)$, decrements *TTL*, and rebroadcasts the *LSU*.

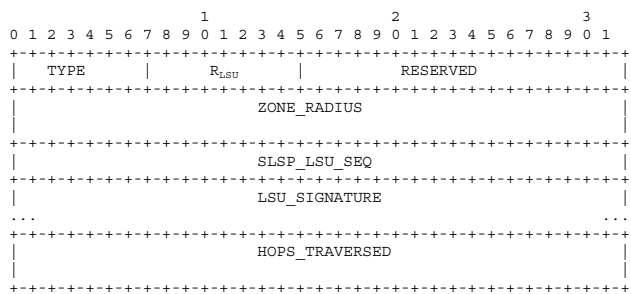

Figure 1: *LSU* Header

The provided information is discarded after a *confirmLS* timeout, unless both nodes incident on a link report the same state. Finally, NLP notifications result in discarding an update relayed by a misbehaved node. The flooding of the *LSU* packets renders the protocol resilient against malicious failures (e.g., packet dropping, alteration, or modification of the packet's *hops_traversed* field). Meanwhile, the localized flooding keeps the transmission and processing overhead low.

---

[3] Hash chains have a wide range of applications; in the MANET context, they have been used to assist in hop count authentication [19].

### 2.5. Public Key Distribution

Nodes use *Public Key Distribution (PKD)* packets, or attach their certified keys to *LSU* packets. *PKD* packets, shown in Figure 2, are flooded throughout the zone, or they may be distributed less frequently throughout an extended zone.

The *LSU*-based key broadcast provides for timely acquisition of the key and thus validation of routing information to nodes that move into a new zone. It also reduces to a great extent the transmission of PKD packets, thus reducing the message complexity. On the other hand, the distribution within an extended zone can reduce the delay of validating new keys when nodes outside a zone eventually enter the zone.

Key broadcasts are timed according to the network conditions and the device characteristics. For example, a node can rebroadcast its key when it detects a substantial change of the topology of its zone; that is, if at least some percentage of nodes has departed from the node's neighborhood since the last key broadcast.

The certificate "vouches" for the public key. Additionally, the authenticity and freshness of the *PKD* packet are verified by a signature from the node that possesses and distributes the key. The *PKD* sequence number is set to the next available value, following the increasing values used for *LSU* packets. When the *LSU*-based key broadcast is used, no additional *PKD* signature is required.

Nodes validate PKD packets only if they are not already aware of the originator's public key. Upon validation, $E_V$ and the corresponding source *IP* address are stored locally, along with the corresponding sequence number.[4] Each node can autonomously decide whether to validate a key broadcast or not. For example, if it communicates with a nearby destination, it might have no incentive to validate a *PKD* that originates from a node a large distance away. Similarly, a validation could be avoided if the node considers its topology view broad enough, or sufficient to support its communication. This could happen for a dense network or zone, when not all physically present links are necessary.

### 2.6. Protection from clogging DoS attacks

In order to guarantee the responsiveness of the routing protocol, nodes maintain a priority ranking of their neighbors according to the rate of queries observed by NLP. The highest priority is assigned to the nodes generating (or relaying) requests with the lowest rate and vice versa. Quanta are allocated proportionally to the priorities and non-serviced, low-priority

---

[4] This information is maintained in a *FIFO* manner. If the entire sequence is covered, a new key is generated and distributed, after the node voluntarily remains "disconnected" for a period equal to NLP's *neighbor_lost*. This temporary disconnection ensures that the possible change of the node's IP address does not cause neighbors to perceive this as a possible attack (i.e., spoofing of an IP address).

queries are eventually discarded. Within each class, queries are serviced in a round-robin manner.

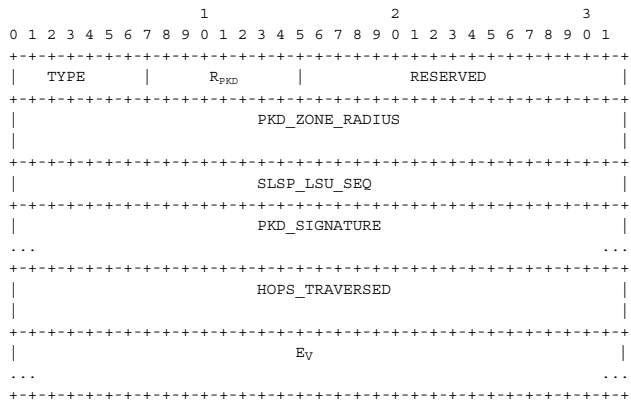

Figure 2: PKD packet

Selfish or malicious nodes that broadcast requests at a very high rate are throttled back, first by their immediate neighbors and then by nodes farther from the source of potential misbehavior. On the other hand, non-malicious queries, that is, queries originating from benign nodes that regulate in a non-selfish manner the rate of their query generation, will be affected only for a period equal to the time it takes to update the priority (weight) assigned to a misbehaving neighbor. In the meantime, the round robin servicing of requests provides the assurance that benign requests will be relayed even amidst a "storm" of malicious or extraneous requests.

Moreover, malicious floods of spurious *PKD* packets are countered by several mechanisms: (i) NLP imposes a bottleneck thanks to the *lost neighbor* timeout, (ii) *PKD* packets will not propagate more than *R* hops, unless they are "carried" farther by adversaries (e.g., when they don't update the *hops_traversed* field), (iii) nodes can autonomously decide whether to validate a public key or not (e.g., for an very high *R*), and (iv) *PKD* packets are also subject to restrictions imposed by the above-mentioned penalizing priority mechanism.

## 3. Discussion

SLSP remains vulnerable to colluding attackers; two malicious nodes $M_1$, $M_2$ may be able to convince nodes in their zones of a non-existent $(M_1, M_2)$ link. However, it is important that any coalition of adversaries can fabricate connectivity only among themselves. Furthermore, the use of a protocol such as SMT on top of SLSP will promptly reveal such forged links, unless the adversaries relay, i.e., tunnel data across such a "virtual" link.

The use of public key cryptography may be a concern as well, especially for resource-constrained devices. Clearly, SLSP nodes should be able to perform public key operations. Since nodes periodically generate (sign) updates and receive (validate) updates more frequently, a cryptosystem with the properties of RSA would be preferable. Most importantly, SLSP provides for a number of ways nodes can minimize their processing while retaining the efficiency of the topology discovery. First, nodes reduce or increase their LSU broadcast period according to the network conditions. With the selection of the appropriate update strategy, a reduced rate of broadcasts does not affect the ability of nodes to maintain up-to-date connectivity information. Moreover, only a small fraction of PKD packets needs to be validated by nodes. Furthermore, the mechanisms that mitigate clogging denial of service attacks ensure that spurious traffic will not consume node resources.

## 4. Conclusions and future work

We proposed a secure link state protocol (SLSP) for mobile ad hoc networks. SLSP is robust against individual Byzantine adversaries. Its secure neighbor discovery and the use of NLP strengthen SLSP against attacks that attempt to exhaust network and node resources. Furthermore, SLSP can operate with minimal or no interactions with a key management entity, while the credentials of only a subset of network nodes are necessary for each node to validate the connectivity information provided by its peers.

The securing of the locally proactive topology discovery process by SLSP can be beneficial for MANET for a number of reasons. The security mechanisms of SLSP can adapt to a wide range of network conditions, and thus retain robustness along with efficiency. As the next step of our research, we will present a detailed performance evaluation of SLSP, both independently and as part of a hybrid framework (i.e., combine it with a secure reactive protocol), and for various network instances and node processing capabilities.

## References

[1] B. Dahill, B.N. Levine, E. Royer, C. Shields. "A Secure Routing Protocol for Ad Hoc Networks." *Technical Report UM-CS-2001-037*, EE&CS, Univ. of Michigan, August 2001.

[2] P. Papadimitratos and Z.J. Haas. "Secure Routing for Mobile Ad Hoc Networks," *SCS Communication Networks and Distributed Systems Modeling and Simulation Conference (CNDS 2002)*, San Antonio, TX, January 27-31, 2002.

[3] Y-C. Hu, A. Perrig, D. B. Johnson. "Ariadne: A Secure On Demand Routing Protocol for Ad Hoc Networks." *MobiCom '02*, Sept. 23-26, Atlanta, GA.

[4] M.R. Pearlman and Z.J. Haas. "Determining the Optimal Configuration of for the Zone Routing Protocol." *IEEE JSAC, special issue on Ad-Hoc Networks,* vol. 17, no.8, Aug. 1999.

[5] Y-C. Hu, D.B. Johnson, and A. Perrig. "Secure efficient distance vector routing in mobile wireless ad hoc networks." *Fourth IEEE Workshop on Mobile Computing Systems and Applications* (WMCSA '02), Jun. 2002.

[6] P. Papadimitratos and Z.J. Haas. "Secure Message Transmission for Mobile Ad Hoc Networks." *Submitted for publication*.

[7] W. Peng and X. Lu. "On the reduction of broadcast redundancy in mobile ad hoc networks." *Proceedings of MOBIHOC '00*, Boston, MA, Aug. 2000.

[8] Z.J. Haas and M.R. Pearlman. "The Performance of Query Control Schemes for the Zone Routing Protocol." *ACM/IEEE Transactions on Networking,* vol. 9, no. 4, pp. 427-438, Aug. 2001.

[9] R. Perlman. "*Interconnections: Bridges and routers.*" Addison Wesley, Reading, MA (Aug 1997).

[10] S. Murphy, et al. "Retrofitting Security into Internet Infrastructure Protocols." *Proceedings of DARPA Information Survivability Conference and Exposition* (DISCEX'00), 2000.

[11] C. Partridge et al. "FIRE: flexible Intra-AS routing environment." *ACM SIGCOMM Computer Comm. Review*, Vol. 30, Issue 4, Aug. 2000.

[12] P. Papadimitratos and Z.J. Haas, "Securing the Internet Routing Infrastructure," *IEEE Communications Magazine*, Vol. 40, No. 10, Oct. 2002.

[13] J. Kong, P. Zerfos, H. Luo, S. Lu and L. Zhang. "Providing Robust and Ubiquitous Security Support for Mobile Ad-Hoc Networks." *IEEE ICNP 2001, Riverside, CA*, Nov. 2001.

[14] J.P. Hubaux, L. Buttyan, and S. Capkun. "The quest for security in mobile ad hoc networks." *2nd MobiHoc*, CA, Oct. 2001.

[15] L. Zhou and Z.J. Haas. "Securing Ad Hoc Networks." *IEEE Network Magazine*, vol. 13, no.6, Nov./Dec. 1999.

[16] M. Hattig, Editor, "Zero-conf IP Host Requirements," *draft-ietf-zeroconf-reqts-09.txt*, IETF MANET Working Group, Aug. 31[st], 2001.

[17] G. Montenegro and C. Canstellucia. "SUCV Identifiers and Addresses." *Draft-montenegro-sucv-02.txt*, work in progress.

[18] NIST, Fed. Inf. Proc. Standards. "Secure Hash Standard." *Pub. 180*, May 1993.

[19] R. Rivest. "The MD5 Message-Digest Algorithm." *RFC 1321*, Apr. 1992.

[18] W. Stevens. "*Unix Network Programming.*" Prentice-Hall.

[19] M. G. Zapata, N. Asokan. "Securing Ad hoc Routing Protocols." *1st ACM WiSe*, Atlanta, GA, Sept. 28, 2002.